# Probing Terahertz Metamaterials with Subwavelength Optical Fibers

**Martin Girard,[1] and Maksim Skorobogatiy[1*]**

[1]*École Polytechnique de Montréal, Engineering Physics Department, C.P. 6079, Centre-Ville Montreal, QC H3C 3A7, Canada*
[*]*maksim.skorobogatiy@polymtl.ca*

**Abstract:** Transmission through a subwavelength terahertz fiber, which is positioned in parallel to a frequency selective surface, is studied using several finite element tools. Both the band diagram technique and the port-based scattering matrix technique are used to explain the nature of various resonances in the fiber transmission spectrum. First, we observe that spectral positions of most of the transmission peaks in the port-based simulation can be related to the positions of Van Hove singularities in the band diagram of a corresponding infinite periodic system. Moreover, spectral shape of most of the features in the fiber transmission spectrum can be explained by superposition of several Fano-type resonances. We also show that center frequencies and bandwidths of these resonances and, as a consequence, spectral shape of the resulting transmission features can be tuned by varying the fiber-metamaterial separation.

## 1. Introduction

Metamaterials have been extensively studied during the last decade. These artificial materials are built to exhibit behavior that is normally not found in nature, with the most notable example being negative refractive index materials. Two dimensional metamaterials or Frequency Selective Surfaces (FSSs) have been demonstrated to exhibit novel behavior such as advanced polarization manipulation [1], beam steering[2], invisibility cloak [3] and high-Q terahertz (THz) frequency selection [4, 5], to name a few. Tuning metamaterials by varying the local or global density of charge carriers [2, 6, 7] has also been demonstrated. One of the enabling geometries that has been thoroughly studied and commonly used in the FSSs is based on the periodic arrays of Split-Ring Resonators (SRRs).

Interrogation of the frequency selective surfaces in THz is mainly performed by sending a THz beam perpendicular to such a surface. Experimentally, interrogation perpendicular to the sample surface is frequently the only option available due to alignment limitations experienced by many THz setups. When interrogating FSS perpendicular to its surface, one is limited with respect to the possible directions of the electric and magnetic fields, which are always located in the plane of a FSS. Despite of all these limitations, many remarkable properties of the FSSs have been demonstrated including coupling between light and dark modes through the use of asymmetry in the metamaterial structure [8].

The optical properties of frequency selective surfaces have been much less studied when the probing light is propagating parallel to the FSS surface. One way to realize such beams in practice is to use optical waveguides suspended parallel to FSS. One advantage of this arrangement is that interaction length between THz beam and FSS can be made arbitrarily long (in the absence of material losses). Additionally, one can control polarization of the probing light to realize either magnetic or electric field to be perpendicular to a FSS.

While waveguide-FSS systems have not been yet detailed, at the same time, the waveguide-resonator systems have been extensively studied (see, for example, [9]) and showed to exhibit various exciting phenomena including tuneable Electromagenic Induced Transparency (EIT)-like transmission [10, 11] and Fano resonances [12]. In fact, it has been shown that EIT-like transmission phenomenon and Fano resonances have essentially the same origin. Namely, transmission spectra corresponding to these phenomena can be understood in terms of the individual properties of several closely spaced (in frequency) resonances using standard considerations of the scattering theory [13]. Moreover, in nanostructures presenting two resonances, EIT-like transmission and superscattering have been shown to be related [14]. It is important to note that many of the modern discussions on the subject of transmission characteristics of the waveguide-resonator systems take roots in the analysis of similar systems in photonic crystals that exhibit many of the same effects [15]. In order to realize EIT and Fano resonances experimentally, one frequently employs single mode waveguides coupled to various resonators such as optical cavities, photonic lattice defects, quantum dots

[16], etc. The EIT phenomenon is typically observed when waveguide is coupled to several slightly detuned resonators. It has also been demonstrated that both the waveguide-resonator coupling strength, as well as phase can be used to tune the EIT resonances [17].

Many potential applications of metamaterials envision their use as point devices. Currently, one assumes that, first, light is delivered by some means to the metamaterial location, then, light interrogates the metamaterial-based device, and finally, transmitted or reflected light is analyzed at the device location. In practical applications, however, it is more convenient to use waveguides to get the light in and out of the point devices. This not only allows a convenient remote excitation and analysis of the transmitted light, but it also allows reliable and, potentially, tuneable coupling of light in and out of a device. A key component of the integrated optics that enables this integration concept is a low-loss, waveguide. Unfortunately, in the THz spectral range, development of the low-loss, low-dispersion waveguides that allow efficient excitation using standard THz sources proved to be a challenging task. As one potential solution, our group has recently demonstrated a variety of subwavelength polymer-based low-loss fibers that can deliver THz power over long distances (meters) [18]. The next natural question for us is whether such fibers can be used to realize a complete THz fiber-based system that enables light delivery, device interrogation, and remote light analysis similarly to the systems developed in the near-IR spectral range [19, 20].

In this work, we study the use of subwavelength terahertz fibers as light delivery and interrogation platform for operation with metamaterial-based point devices. Particularly, we study subwavelength fiber that is placed parallel to the frequency selective surface. Thus defined coupler mostly operates in a strong coupling regime as evanescent fields of the subwavelength fiber show strong presence outside of the fiber core. We note also that different polarization states can be realized in this arrangement with electric field either parallel or perpendicular to the metamaterial surface depending on polarization of the fiber mode. Additionally, given the large sizes of THz fiber and split ring resonators (hundreds of microns), a tunable coupling between metamaterial and fiber can be easily achieved by mechanical actuation, thus opening a way for active device applications such as frequency scanning, switching and dynamic filtering. To the best of our knowledge, it is the first time that such subwavelength fiber-metamaterial couplers are studied.

The paper is organized as follows. First, we detail geometry of the fiber-FSS coupler considered in this work, and then we detail numerical approaches used in our simulations. Second, we summarize predictions of a scattering theory for the forms of transmission curves in the presence of several resonances. These considerations will be used in the following chapters to explain the results of our simulations. Third, for an infinite fiber-FSS structure we present the corresponding band diagram calculated using a supercell approximation. Fourth, we study transmission through the fiber suspended over a single period (along the fiber length) of a FSS. Particularly we investigate the effect of supercell transverse size and material losses on the fiber transmission spectrum. Fifth, using a particular supercell we show that spectral properties of the resonant features in the fiber transmission spectrum are tunable by varying the fiber-FSS distance. Changes in the resonant features as a function of the fiber-FSS distance are then interpreted quantitatively using the line shapes from the scattering theory. Finally, we investigate convergence in the fiber transmission properties when increasing the number of metamaterial periods along the fiber length. In this regime, we see a good correspondence between results of the transmission calculations and predictions of the band diagram calculations. Particularly, positions of the Van Hove singularities, as detected from the structure of a band diagram, correspond exactly to the positions of resonances in the transmission calculations. Additionally, we observe convergence in the peak spectral widths when increasing the number of metamaterial periods.

## 2. Geometry of a fiber-metamaterial coupler

In this work we perform two types of calculations. First, we compute the band diagram corresponding to an infinite periodic system consisting of a fiber suspended in parallel to the metamaterial. Second, we conduct scattering matrix transmission simulations with two ports, where we assume a finite number of periods in a coupler. Schematic of a supercell used in our simulations in presented Fig. 1. There, the air-clad fiber has radius $R = 200$ μm and it is made of a polymer with refractive index $n = 1.55$. The fiber is suspended over metamaterial at a distance $H$. The metamaterial is build on a $700$ μm – thick fused silica substrate with refractive index $n = 1.966$. The substrate is patterned with split ring resonators made of perfect electrical conductor with height $h=50$ μm. The period of metamaterial cell is $\Lambda = 400$ μm. Finally, for each simulation presented in this work we specify the number of SRRs in the transverse direction $N_t$, as well as the number of SRRs in the longitudinal direction $N_l$.

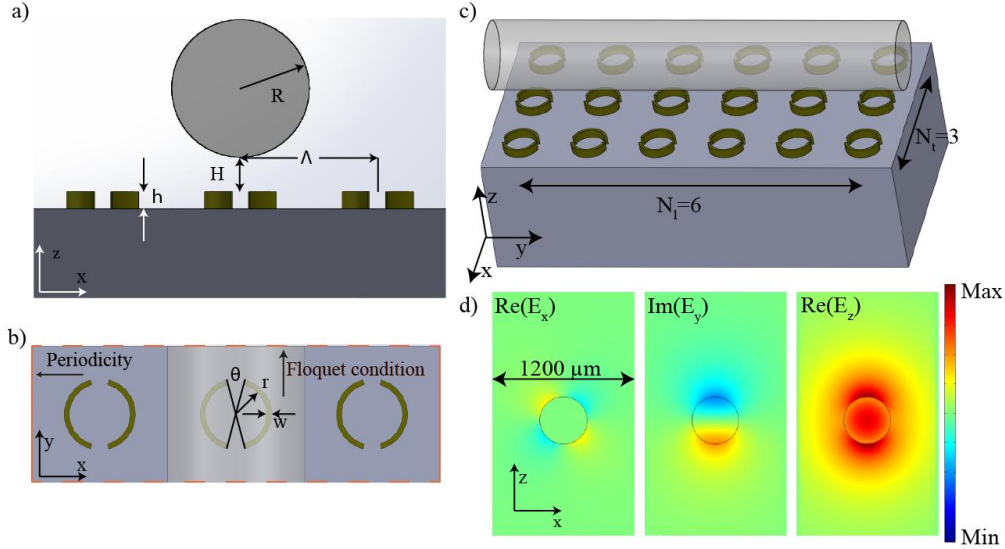

Fig. 1. Schematic of a unit cell used in simulations. a) side view, b) top view. Unless stated otherwise $R = 200$ μm, $h =50$ μm, $\Lambda = 400$ μm, $\theta = 30°$, $w = 15$ μm and $r = 90$ μm. c) 3D rendering of a fiber-FSS system featuring 3 transverse and 6 longitudinal periods. d) Linearly polarized (along z) fundamental $HE_{11}$ mode of a single mode fiber is used as a port condition.

COMSOL finite element software is used in all our calculations. The direction of wave propagation is along the $y$ axis. Along this axis we either use the Floquet boundary condition for the band diagram calculation, or the port boundary condition for the transmission calculations. In the direction normal to the plane of FSS ($z$ axis) a $6000$ μm air gap is added above the substrate as well as $2000$ μm air gap is added below the substrate. In $z$ direction, the unit cell is terminated with the PML boundary conditions. The periodic boundary condition is applied in the transverse $x$ direction, in order to simulate an infinitely wide metamaterial.

In the case of transmission calculations we use the port boundary conditions. Both at the input port (a source waveguide) and at the output port (an outgoing waveguide) we take into consideration only one guided mode which is the fundamental mode of a single-mode fiber polarized in the direction perpendicular to the metamaterial surface (the leading component of the modal electric field is $E_z$, see Fig. 1(d)). As a consequence, the minor $E_x$ component of the modal electric field is antisymmetric in the $x$ direction, which, in principal, can result in conflicting boundary conditions at the supercell boundary where we impose periodic boundary conditions. To prevent this from happening, the $HE_{11}$ modal fields (that constitute port boundary conditions) are somewhat modified to go to zero at the periodic boundaries.

We note that a certain caution has to exercised when using periodic boundary conditions. Particularly, the width of the supercell has to be wide enough to prevent significant overlap of the fiber mode with that of an image fiber created by the periodic boundary conditions. In our simulations we observe that periodic boundary conditions do not affect strongly the system response at higher frequencies ($f>250$ GHz) when at least 3 SRRs are used in the transverse direction. At lower frequencies, transverse modal fields of the fundamental mode of a subwavelength fiber show very slow logarithmic decay [21] outside of the fiber core. Therefore, at such frequencies computational artifacts could be introduced due to coupling between the subwavelength fiber and its images.

Finally, we note that in practical implementations of the THz metamaterials the height $h$ of the split ring resonators is much smaller than the one used in our simulations, and it is typically on the order of several microns. However, using such a small height is impractical even if finite element solver is used as such thin layers result in the intractable number of elements. For example, when performing transmission calculation using a unit cell consisting of only a single row of 3 SRRs, $50 \cdot 10^3$ elements are generated (6.4 GB) when $h = 50~\mu m$, whereas $135 \cdot 10^3$ elements (17 GB) are generated when $h = 5~\mu m$. A major problem arises when modeling longer systems. Thus, for $h = 50~\mu m$ and a 10 period-long supercell (total of 30 SRRs), 70 GB of memory and 40 minutes are required per single frequency calculation. Therefore, with 128 GB of memory and $h = 50~\mu m$ we are limited by 11 periods (33 SRRs).

## 3. Band diagram of the fiber-metamaterial coupler

First, we present the results of band diagram calculations. In Fig. 2 we show dispersion relations of the modes of an infinitely periodic fiber-metamaterial coupler (periodicity along the fiber direction). The unit cell used in the simulations (see Fig. 1(a)) features three SRR in the transwerse direction $N_t=3$, and a single SRR in the longitudinal direction. Fiber-metamaterial separation is $h=50~\mu m$. Floquet boundary conditions are used at the supercell boundaries terminating the fiber. The band diagram of an infinitely periodic fiber-metamaterial system is presented in Fig. 2(b). To show the effect of split ring resonators, we have repeated simulation by keeping exactly the same geometry, however, without the SRRs on the slab surface (see Fig. 2(a)). For the ease of comparison, resultant band diagram is presented within the same first Brillouin zone as in the case of a fiber-metamaterial system. Color code reflects the average value of the norm of a $y$ component of a Poynting vector taken over the volume of a fiber to the root mean square of the electric field in the whole computational cell. Thus, the blue color corresponds to low fraction of the modal fields in the fiber, while the red color corresponds to strong presence of the modal fields in the fiber.

As it is well know from the theory of circular step-index fibers, their fundamental $HE_{11}$ mode is doubly degenerate. When bringing a planar silica substrate into the fiber vicinity, this degeneracy is lifted and the fiber modes can then be characterized as predominantly $x$ or $z$ polarized depending on the leading transverse component of their electric fields. The perturbation seen by the subwavelength fiber is very strong due to a significant presence of its fields in the metamaterial substrate region. The fundamental $HE_{11}$ mode of an unperturbed fiber is, thus, absent from Fig. 2 and is replaced by the two slab-fiber supermodes. Moreover, fiber modes and slab modes can show strong hybridization in the vicinity of phase matching points, which results in standard avoiding crossing behavior of the modal dispersion relations (see a part of Fig. 2(a) within a large circle). For small fiber-slab separations, coupling between the modes of a fiber and a slab can be very strong, thus resulting in significant changes in the curvature of the dispersion relations of the hybrid modes in the vicinity of a phase matching point. This, in turn, leads to the high values of the group velocity dispersion of the hybrid modes. It is important to note from Fig. 2(a) that accidental crossing of the fiber and slab modes does not necessarily lead to avoiding crossing phenomenon or modal hybridization. In fact, when symmetries of the fiber and slab modes are not compatible this results in zero coupling strength, thus, no hybridization between the two modes occurs.

Addition of the SRRs brings two important modifications to the structure of the band diagram. The first one is manifested by the SRR-induced interactions between otherwise non-interacting fiber and slab modes or the two slab modes characterized by the incompatible symmetries. The effect of such an induced interaction is most pronounces when dispersion relation of one of the modes is folded back into the first Brillouin zone (see four regions in Fig. 2(a),(b) within small circles). In this case, at a phase matching point the slopes of the two dispersion relations have opposite signs, and, therefore, the result of the interaction is in the opening of a local bandgap.

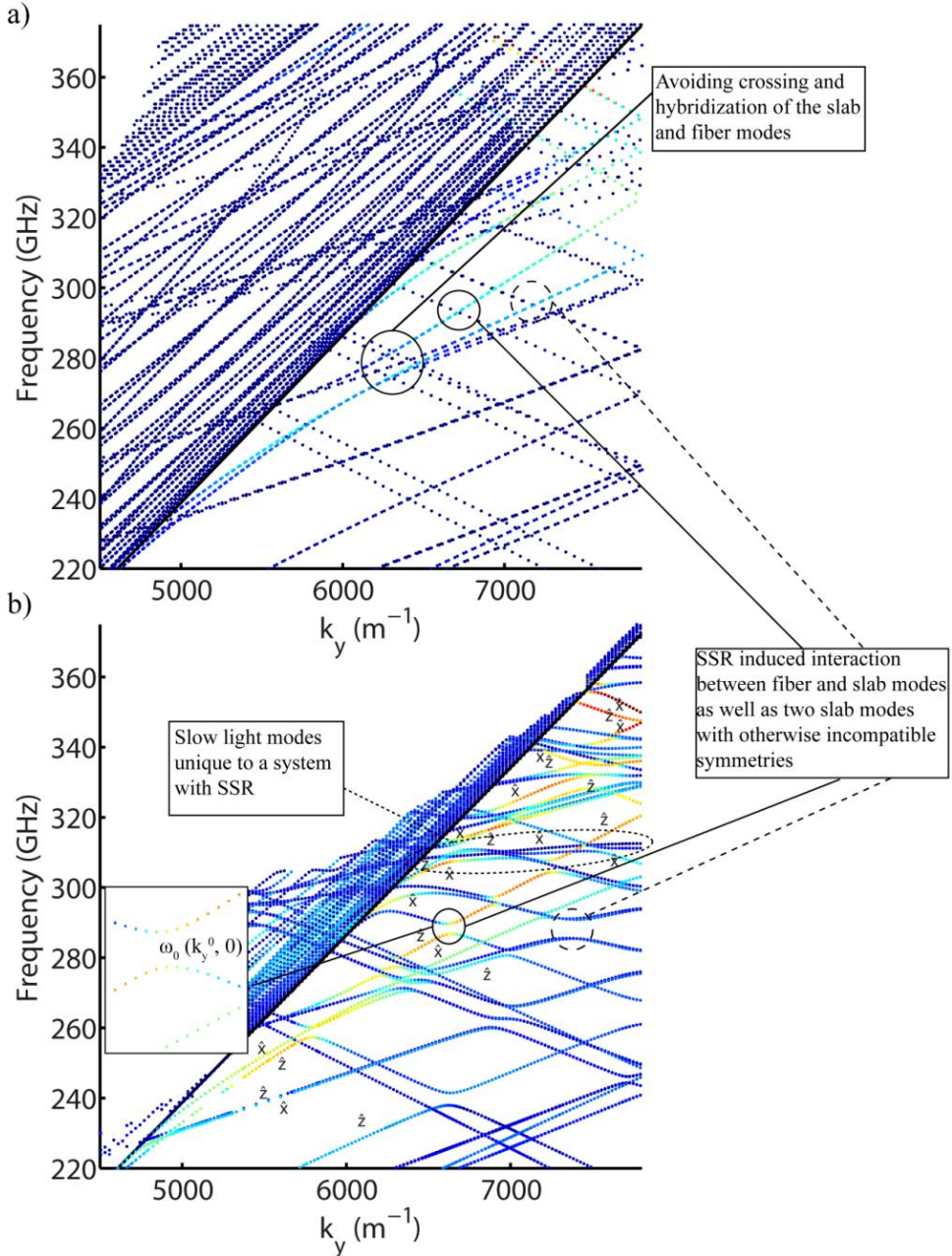

Fig. 2. Band diagrams for a) fiber-slab system without SRRs, b) fiber-metamaterial system with SRRs. Polarization of some of the modes is indicated as *x*, or *z* depending on the leading component of the electric field. Color for each optical state indicates the degree of field localization in the fiber core region.

Note that due to periodic boundary conditions in the transverse direction, the band diagrams presented at Fig. 2 are calculated along the line of high symmetry defined by $k_x=0$. This means that in the vicinity of the edges of local bandgaps, the full dispersion relation of the allowed optical states is of either a parabolic or hyperbolic form $\omega = \omega_0 + A \cdot k_y^2 + B \cdot (k_x - k_x^0)^2$. Consequently, at frequencies corresponding to the edges of the local bandgaps (see insert of Fig. 2(b)) one expects appearance of the Van Hoff singularities in the local density of states (see chapter 6.4 in [6], for example). As we will show in the following sections, spectral position of the resonant features in the fiber transmission spectrum coincide with positions of the Van Hoff singularities as determined from the fiber-metamaterial band diagram. This establishes the fundamental relation between the fiber transmission and band diagram calculations. The second impact of the introduction of SRRs is appearance of slow light bands characterized by flat dispersion relations (region of Fig. 2(b) within the dashed ellipse).

## 4. Scattering theory

In the remaining sections we perform fiber transmission simulations using port boundary conditions. In our interpretation of the resonant features in the fiber transmission spectrum we use scattering theory formulation presented in [13], as well as results of the band diagram calculations presented in the previous section. The scattering theory considers interaction of a waveguide guided mode with a certain number of discrete resonator states. In our case, guided mode is the fundamental mode of a subwavelength fiber used by the port boundary conditions. At the same time, the discrete resonator states are the hybridized modes of a fiber-metamaterial system in the vicinity of Van Hove singularities. Note from Fig. 2(b) that Van Hove singularities usually come in pairs as they are the result of anticrossing between two modal dispersion relations with different slope signs. Therefore, in the transmission spectrum we expect to see various line shapes that are formed by a pair of resonances.

In the lossless system, scattering theory predicts that frequency dependent transmission coefficient trough the fiber $t(f)$ can be cast in terms of the frequency dependent total scattering crossection $\sigma_t(f)$ as:

$$|t|^2(f) = \frac{\sigma_t(f)^2}{1+\sigma_t(f)^2}. \quad (1)$$

In turn, $\sigma_t(f)$ can be calculated using contributions from the scattering crossections $\sigma_i(f)$ of the individual resonances:

$$\sigma_t(f)^{-1} = \sum_j \sigma_j(f)^{-1}, \quad (2)$$

which in the lossless case are:

$$\sigma_j(f) = (f - f_j)/\gamma_j, \quad (3)$$

where $f_j$ is a resonance center frequency, and $\gamma_j$ is the coupling strength between the resonator mode and a guided mode of a waveguide. In our case, resonant frequencies are determined by spectral positions of the Van Hove singularities. In the limit of weak coupling between the fiber and metamaterials modes, Van Hove singularities are found in the region of avoiding crossing between dispersion relation of a fiber mode and dispersion relation of a particular slab mode that was folded back into the first Brillouin zone due to its interaction with a periodic SRR array.

Expressions (1)-(3) are found by using a lossless transfer matrix formulation that is a result of solution of a system of coupled differential equations that describe changes in the time dependent intensities of the localized resonator states due to their coupling (leakage) into the guided state of a waveguide. In a more general case when resonator states are lossy, solution for the frequency dependent transmission $t(f)$ and reflection $r(f)$ coefficients can be found by resolving a generalized transfer matrix formulation after introduction of the loss parameter $\Gamma$, which is the rate of decay of the resonator state to absorption and far field radiation. Particularly, for a system of $N_r$ localized lossy resonator states coupled to a guided mode, the transfer matrix method [22] requires resolution of the following equation:

$$\begin{pmatrix} t(f) \\ 0 \end{pmatrix} = M(f) \begin{pmatrix} 1 \\ r(f) \end{pmatrix}, \quad (4)$$

where transfer matrix $M(f)$ is defined as:

$$T = \prod_{j=1}^{N_r} \begin{pmatrix} 1 - \dfrac{i\gamma_j}{f - f_j + i\Gamma_j} & -\dfrac{i\gamma_j}{f - f_j + i\Gamma_j} \\ \dfrac{i\gamma_j}{f - f_j + i\Gamma_j} & 1 + \dfrac{i\gamma_j}{f - f_j + i\Gamma_j} \end{pmatrix}. \quad (5)$$

For a single lossy resonant state coupled to a guided fiber mode, power transmission coefficient and total losses can be calculated by solving (4) using (5) to find:

$$|t|^2(f) = \frac{(f - f_0)^2 + \Gamma^2}{(f - f_0)^2 + (\gamma + \Gamma)^2}, \quad (6)$$

which defines a resonant dip in the transmission spectrum. In the lossless limit $\Gamma = 0$ the dip is complete and transmission is exactly zero at the resonant frequency. From expression (6) it also follows that in the immediate vicinity of the resonance frequency $|f - f_0| \ll \gamma$, and in the limit of low loss (or strong coupling) $\Gamma \ll \gamma$, transmission through the waveguide is:

$$|t|^2_{\min} \approx \Gamma^2 / \gamma^2. \quad (7)$$

We will use this important result later in the paper to explain changes in the transmission spectrum in the presence of losses.

Another important case is the interaction between two resonances that lead to appearance of several well-known line shapes such as EIT-like resonances and non-symmetric Fano line shapes, which we observe in the fiber-FSS transmission spectrum. The abovementioned line shapes manifest themselves as one or two spectral dips that are positioned closely to a single transmission peak. For completeness of presentation, the general form of the transmission coefficient in the presence of two lossy resonances can be found from (4) and (5) to be:

$$t(f) = \frac{(f - f_1 + i\Gamma_1)(f - f_2 + i\Gamma_2)}{(f - f_2 + i(\Gamma_2 + \gamma_2))(f - f_1 + i(\Gamma_1 + \gamma_1)) + \gamma_1 \gamma_2}. \quad (8)$$

There are two particular cases that we consider in more details. Firstly, we discuss EIT-like line shape resulting when two almost degenerate (in frequency) resonator states are coupled to the guided fiber mode. In this case, one observes two transmission dips and a transmission peak in the middle between the dips. Secondly, in the case of two well separated resonances, when one of the resonances is broad, and the other one is narrow ($|f_1 - f_2| \sim \gamma_1, \gamma_2 \ll \gamma_1$), one observes a non-symmetric Fano line shape characterized by a transmission dip followed by the transmission peak. The corresponding physical phenomenon where such transmission spectrum is frequently observed is when a localized state (in our case a particular metamaterial-bound mode phase matched with a fiber mode) is competing with a continuum of states (all the other metamaterial-bound modes that are not phase matched at resonance with a fiber mode).

In a general case, for a system of two resonances, in the limit of low loss (or strong coupling) $\Gamma_1 \ll \gamma_1$, $\Gamma_2 \ll \gamma_2$ the maximal value (peak value) of the transmission coefficient can be derived from (4),(5) to be:

$$|t|^2_{\max} \approx 1 - 2\frac{(\gamma_1 + \gamma_2)^2}{(f_1 - f_2)^2}\left(\frac{\Gamma_1}{\gamma_1} + \frac{\Gamma_2}{\gamma_2}\right). \quad (9)$$

At the same time, one can show that transmission amplitudes corresponding to the two dips are similar to those given by expression (7), namely:

$$|t|^2_{min,1,2} \approx \Gamma^2_{1,2}/\gamma^2_{1,2}. \quad (10)$$

From equations (7) and (9), we can deduce two interesting properties of the dips (minima) and peaks (maxima) in the fiber transmission spectrum. Firstly, when introducing small losses into a system, $\Gamma \ll \gamma$, transmission peaks will be lost first before the dips. This is a simple manifestation of the fact that decrease of the peak amplitude (see eq. (9)) is linear with a small parameter $\Gamma/\gamma$, while increase in the dip amplitude is much slower, and it is, in fact, quadratic (see eq. (7)) with the same small parameter. Secondly, when increasing losses, high-Q features (resonances with smaller $\gamma$ values) will be lost first. This is because both dips (7) and peaks (9) are polynomial functions of a small parameter $\Gamma/\gamma$. Therefore, for the same value of loss $\Gamma$, the value of a small parameter will be larger for resonances with higher Q factors. As follows from (7) and (9), this also means that peak and dip amplitudes corresponding to higher-Q features will disappear faster than those corresponding to the lower-Q features. Note that to arrive to this conclusion we have to assume in (9) that the spacing between two resonances is comparable or larger than their total bandwidth $|f_1 - f_2| \sim (\gamma_1 + \gamma_2)$, which is actually always the case in our simulations.

Finally, we would like to talk more about Fano line shapes as these are the most frequently encountered spectral features in our simulations. Fano line shapes are typically obtained in a two resonance system where one resonance (characterized by $\gamma_1$) is broad (to represent a continuum of states), and the other one (characterized by $\gamma_2$) is narrow (to represent a discrete state). We also assume that the narrow resonance is placed within the bandwidth of a broad resonance so that $|f_1 - f_2| \sim (\gamma_1 + \gamma_2)$. In this case, in the vicinity of a narrow resonance, one typically describes the Fano line shape as:

$$|t|^2(f) = \frac{1}{1+q^2} \frac{(q\Gamma_\Phi + f - f_\Phi)^2}{\Gamma_\Phi^2 + (f - f_\Phi)^2}, \quad (11)$$

where $q$ is the Fano asymmetry parameter, $f_\Phi$ is the Fano resonant frequency, and $\Gamma_\Phi$ is Fano width. Assuming lossless resonances $\Gamma_{1,2} = 0$, we can expend (8) in power series around the point of minimal transmission ($f = f_2$) and match (8) and (11) up to the third order in $(f - f_2)$ and $(f - f_\Phi + q\Gamma_\Phi)$, yielding the following relations between the Fano parameters and the parameters of the individual resonances:

$$q = \frac{\gamma_1}{f_1 - f_2}, \quad \Gamma_\Phi = \frac{\gamma_2}{1+q^2}, \quad f_\Phi = f_2 + q\Gamma_\Phi. \quad (12)$$

Note that the frequencies of the minimal and maximal transmissions are respectively $f_{min} = f_2$ and $f_{max} = f_2 + (q + q^{-1})\Gamma_\Phi$.

## 5. Convergence of a supercell approximation

Before we present main results of the transmission calculations using port boundary conditions, we would like to comment on the accuracy of the supercell approximation used in our simulations. In the rest of this section we vary the number of SRRs in the transverse direction (perpendicular to the fiber) $N_t=[3,9,15]$, while using only one SRR along the fiber length $N_l=1$ (see Fig. 3(a)). Transmission through various fiber-FSS supercells are shown in Fig. 3(b), where curves of different color correspond to different number of SRRs in the transverse direction. When comparing the transmission curves for supercells containing 3 and

9 SRRs we note that while many of the broader resonant features are present in both curves, their spectral positions are somewhat different. Moreover, the number of narrow peaks for a wider supercell is considerably larger than the number of peaks for a narrower supercell. When further increasing the width of a supercell to $N_t=15$ (see Fig. 3(c)), one observes that position and shape of many broader peaks do not change, thus indicating convergence of these spectral features, while at the same time, many more narrow peaks appear in the spectrum.

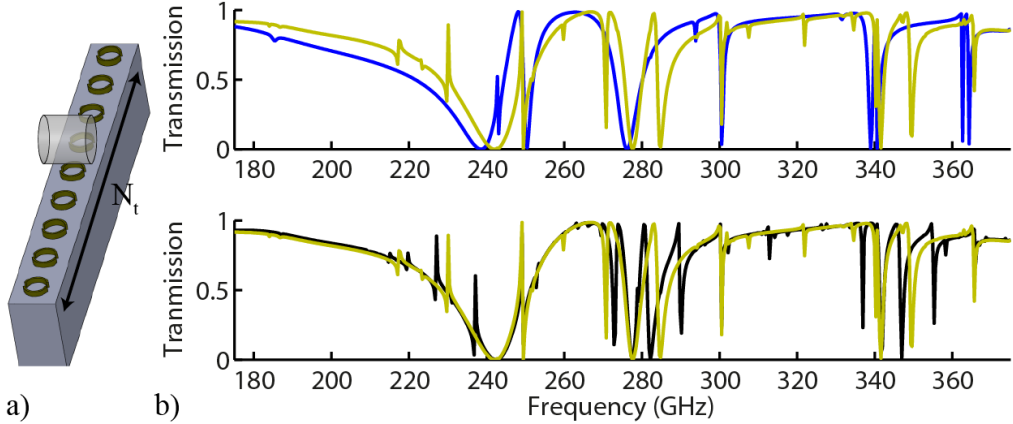

Fig. 3. (a) 3D rendering of a supercell used in simulation with $N_t=9$ SRRs in the transverse direction and a single period in the longitudinal direction. (b) Transmission through supercells of different width, $N_t=3$ (blue), $N_t=9$ (yellow) and $N_t=15$ (black)

We note that, generally, the supercell approximation is most effective when studying the defect states or localized modes that have rapidly decaying fields in the direction of the supercell boundary. Ideally, such localized states have to be located inside of a spectral bandgap so they can be clearly distinguished from the continuum of delocalized states. In our case, however, the defect modes (the ones with a considerable power guided in the fiber) are, in fact, strongly hybridized with the metamaterial modes that are extended over the whole supercell in the transverse direction. When increasing the number of SRRs in the transverse direction, a periodic structure is formed in that direction with a period equal to the width of a supercell. Due to periodic boundary conditions, all the states with transverse wavevectors $k_y \sim 2\pi n/(\Lambda \cdot N_t)$, $n \in Integers$ will contribute to the transmission spectrum. Therefore, the wider is the supercell, the more delocalized transverse modes will contribute to the transmission spectrum. This can be clearly seen from Fig. 3(b) when counting the number of resonance peaks for supercells with sizes $N_t=3,9,15$. From this data it is clear that the number of resonant peaks is proportional to the supercell size. Based on these observations, one would wonder in what sense do these transmission results converge when increasing the width of a supercell, or how do these results relate to experimental measurements.

In fact, convergence of a supercell approximation is easy to observe when introducing material losses into the system. In this case, the hybridized fiber-metamaterial modes become evanescent in the transverse direction and, therefore, their properties converge when the supercell size becomes larger than the transverse decay length of the modal fields. Moreover, in the presence of loss, no new resonant features can form in the transmission spectra when increasing the supercell size beyond a certain value. This is because resonances are formed due to constructive interference of the two counter propagating transverse modes of a metamaterial. Therefore, if the supercell size becomes wider than a characteristic transverse decay length of a metamaterial mode, then no interference is possible, therefore, no new resonant feature can form in the transmission spectrum.

To validate these conclusions, we compare in Fig. 4 the transmission spectra calculated for the supercells of two different widths $N_t=9$ and $N_t=15$ in the presence of losses. First, we note that in the absence of loss (Fig. 4(a)), while there is an overall correspondence between the positions and shapes of the wider resonances, there is clearly no such correspondence for narrow resonances. When introducing material losses for the substrate material into simulations (dashed curves) we observe that sharp peaks rapidly disappear when increasing material losses, while the positions and the widths of the broader peaks remain the same independently of the supercell size. This disappearance of the sharp features is consistent with our predictions from the classic scattering theory (see eq. (9), and discussions of section 4). In our simulations in Fig. 4 we have used 0.02i (Fig. 4(b)) and 0.2i (Fig. 4(a)) for imaginary parts of the substrate permittivity, which correspond approximately to 1 cm$^{-1}$ and 10 cm$^{-1}$ bulk material losses of the fused silica and other typical glasses.

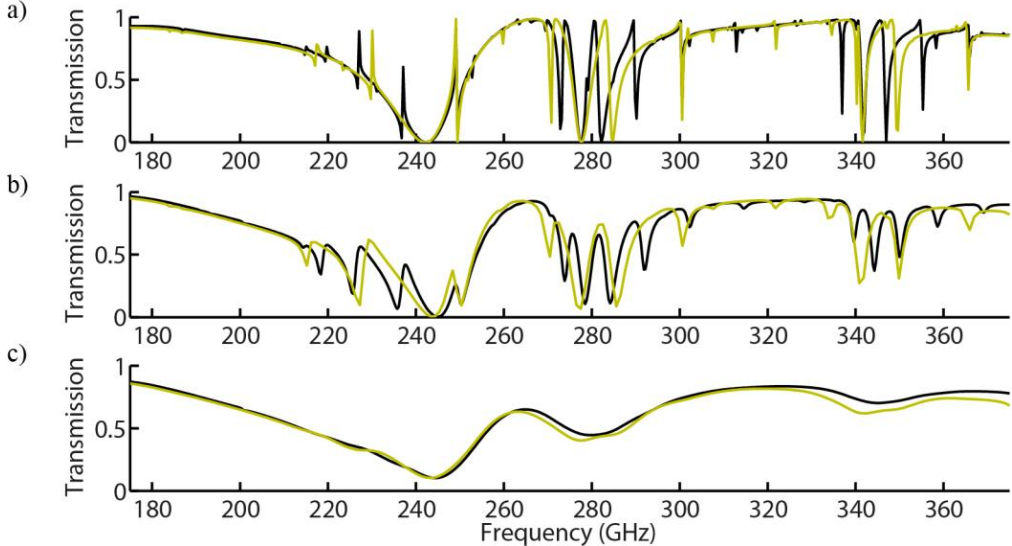

Fig. 4: Transmission through supercells of different width $N_t=9$ (yellow) and $N_t=15$ (black) for different values of the substrate material loss (a) no loss, (b) $Im(\varepsilon)=0.02i$, and (c) $Im(\varepsilon)=0.2i$.

## 6. Effect of the fiber-metamaterial separation on transmission spectrum

We now study the effect of fiber-metmaterial separation on the fiber transmission spectrum. By changing fiber-metmaterial separation one controls the coupling strength between the fiber mode and the metamaterial-bound modes. When using standard Coupled Mode Theory (CMT) [23], one finds that the coupling strength between two modes ($\gamma$ parameter in section 4) is proportional to the overlap integral between their fields. Moreover, the stronger is the coupling strength between the modes, the wider will be the corresponding resonance in the fiber transmission sprectrum. Additionally, in the case of strong coupling, spectral position of the resonant peak becomes dependent on the coupling strength. This is known as coupling-induced frequency shift (CIFS), which had been previously reported in the waveguide-resonator systems [24]. It is, therefore, expected that the resonant features in the fiber transmission spectrum could be tuned by changing the fiber-metamaterial distance $H$.

To demonstrate the impact of fiber-metamaterial separation on the shape and position of the resonances, in Fig. 5 we present several fiber transmission spectra calculated for different values of $H$ in the range of *10-130μm*. The system under study contains $N_t=3$ SRRs in the transverse direction, and one row of SRRs in the longitudinal direction $N_l=1$. In these calculations, all the materials were considered as lossless. Consider in particular two broad resonances at *~238 GHz* and *~250 GHz* that correspond to two metamaterial modes that are strongly coupled to the fiber mode. As expected, the resonances are broader for smaller fiber-metmaterial separations due to enhanced coupling. Moreover, for these two resonances, their resonant frequencies (frequencies of zero transmission) depend strongly on $H$ due to coupling-induced frequency shift effect. This is especially pronounced for a wider resonance at *~238

*GHz* as the coupling strength for this resonance in the strongest among all the resonances presented in Fig. 5. This finding is consistent with the study of CIFS reported in [24], where the authors have concluded that the shift in the resonance position is inversely proportional to the Q-factor of the resonance.

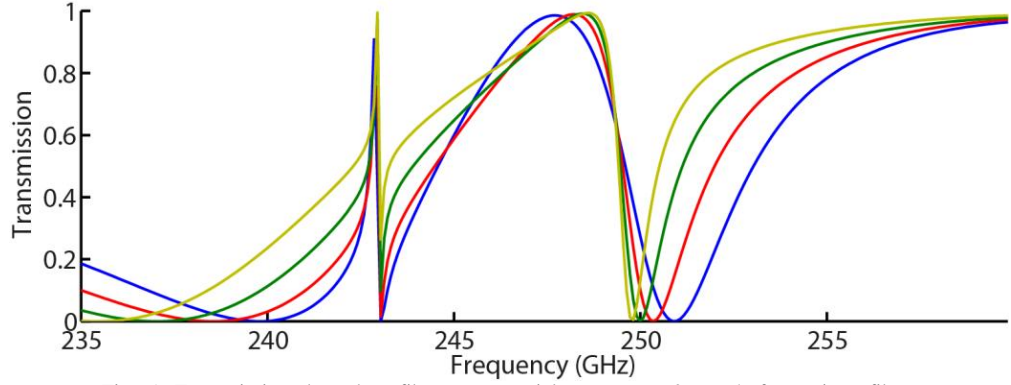

Fig. 5. Transmission through a fiber-metamaterial system $N_r=3$, $N_l=1$, for various fiber-metamaterial separations $H = 10\mu m$ (blue), $H = 50~\mu m$ (red), $H = 90~\mu m$ (green) and $H = 130~\mu m$ (yellow).

We now consider in more details, how the resonant position and the resonant width change with the fiber-metamaterial distance. As we have noted earlier, in the weak coupling limit, resonances in the fiber transmission spectrum are typically observed at the phase matching point between the fiber mode and a particular matamaterial mode. Resonance width in this case is proportional to the coupling strength between the two modes, while the coupling strength is proportional to the overlap integral between two interacting modes [23]. The fields of a subwavelength fiber mode show logarithmic decay outside the fiber, while the metamaterial modes are well confined and show exponential decay outside of metamaterial. Therefore, one expects that overlap integral, and, consequently, coupling strength should decrease exponentially fast with *H*. This means that the resonance position $f_r$ (point of zero transmission) and the resonance width $\gamma_r$, could be fitted as follows:

$$f_r(H) = f_0 + \Delta f_r \cdot \exp(-H/H_f)$$
$$\gamma_r(H) = \Delta\gamma_r \cdot \exp(-H/H_\gamma)$$
, (13)

where $f_0$ the resonant frequency in the limit of zero coupling, $\Delta f_r$ is the tuning range of the peak position, $\Delta\gamma_r$ is the tuning range of the resonance width, while $H_f$ and $H_\gamma$ are the characteristic fiber-metamaterial separations required for tuning.

As an example, in Fig. 6(a) we study dependence of the transmission spectrum corresponding to the resonance located at ~*240 GHz* on the values of fiber-metamaterial separation *H*. Position of the resonant frequency $f_r(H)$ and peak bandwidth $\gamma_r(H)$ are extracted from the transmission spectra by fitting them with (4), (5), while using three lossless resonances $\Gamma_{1,2,3} = 0$. Resultant values for the peak bandwidth and the peak position (presented as circles in Fig. 6(b)) are then fitted with the analytical dependencies (13) to give $f_0 = 215~GHz$, $\Delta f_r = 25~GHz$, $H_f = 555~\mu m$ for the resonance position, and $\Delta\gamma_r = 13.5~GHz$ $H_\gamma = 838~\mu m$ for the resonance width. Overall, the agreement between numerical results and analytical fit is excellent, except for the values of the peak width in the limit of small fiber-metamaterial separations $H < 125~\mu m$. In fact, at these small values of separation, the broad peak shifts too close to the sharp peak at *243 GHz*. As a consequence, there is significant interaction between the two resonant modes and, as a consequence, simple model (13) becomes inadequate in this regime.

It is also interesting to comment on the modal field distribution at the frequencies of the minimal and maximal transmission. Particularly, in Figs. 6(c), (d), we present the logarithm of the $|E_z|$ field distribution in the *y-z* plane for a separation distance of $H = 50~\mu m$ at the frequencies of minimal transmission (*f = 238.4 GHz*) and maximal transmission (*f = 242.9 GHz*). The imaging plain goes through the fiber center, therefore, in Figs 6 (c), (d) we see the fiber core (on the top) suspended over the metamaterial substrate (in the middle of the figure) with a single SRR positioned between the two. It is clear from Fig. 6(c) that in the case of zero transmission, the power lunched into the fiber through the input port (left side of the figure) is transferred completely into the metamaterial substrate by the time it arrives to the output port (right side of the figure). The field in the metamaterial substrate has a very small overlap with the mode of the output port, therefore, one detects minimum in the fiber transmission. On the other hand, in the case of a perfect transmission (see Fig. 6(d)), the power launched into the fiber at the input port is first transferred into the metamaterial substrate and then back into the fiber by the time it arrives to the output port, therefore, one detects maximum in the fiber transmission.

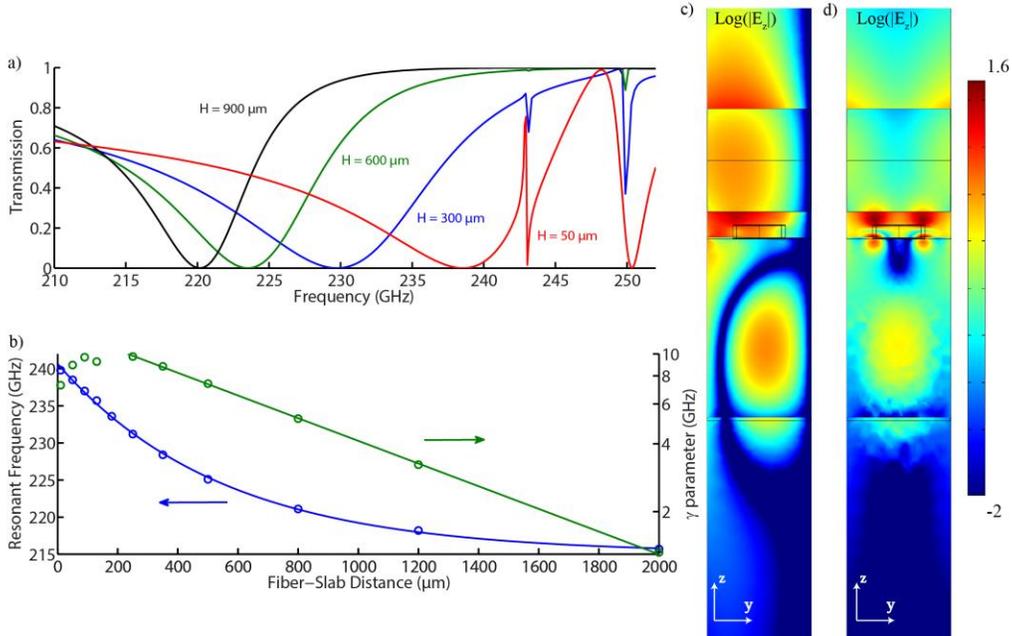

Fig. 6. (a) Changes in the transmission peak located at ~*240 GHz* as a function of the fiber-metamaterial separation *H*. (b) Changes in the peak position (zero transmission) and peak width are fitted very well with exponential dependence on *H*. (c) Electric field distribution at the frequency of zero transmission, and (d) at the frequency of maximal transmission.

## 7. Effect of the fiber-material coupling length on transmission spectrum

In this section we investigate changes in the transmission spectra of a fiber-metamaterial system when changing the number of SRRs along the fiber direction. Particularly, we consider metamaterial containing $N_t=3$ SRRs in the transverse direction, while in the longitudinal direction we vary the number of SRR periods from $N_l=1$ to $N_l=11$. When increasing the number of SRR periods, the response of a finite-size metamaterial should eventually converge to that of an infinitely long periodic system. Therefore, we expect that in the limit or large $N_l$, there should be a direct correspondence between the band diagram structure (see Fig. 2) and the transmission spectra.

In Fig. 7 we present transmission spectra of a long fiber-metamaterial system with $N_l=10$ periods. This spectrum is calculated with resolution of *0.3 GHz*. In the same figure we mark (dotted lines) the spectral positions of various Van Hove singularities as found from the band diagram of the corresponding infinite system (see Fig. 2). To remind the reader, Van Hove singularities are found at frequencies at which dispersion relations of the optical bands show local maxima or minima, and as a consequence, optical density of states at such singularities have particularly high values. From Fig. 7 we note that a great majority of the resonant peaks in the transmission spectrum correspond to Van Hove singularities in the optical density of states. However, as seen from Fig. 7, not all Van Hove singularities found in Fig. 2 manifest themselves as transmission peaks. This is because optical modes at such singularities are either incompatible by symmetry with the fundamental fiber mode (port mode), or because field overlap between the fiber and metamaterial modes is too small.

Efficient coupling between the fiber mode (port mode) and metamaterial modes are observed in the spectral regions where dispersion relation of the fiber mode exhibits avoiding crossing with the dispersion relations of backward propagating slab modes. Backward propagating slab modes are the ones with dispersion relations that are folded back into the first Brulloin zone due to presence of a periodic SRR array (see, for example, a circled region at ~0.29 THz in Fig. 2(b)). In this case, avoiding crossing between fiber and metamaterial modes necessarily results in the creation of a Van Hove singularity, and the resultant hybrid modes in the vicinity of such a singularity show strong presence in the fiber core.

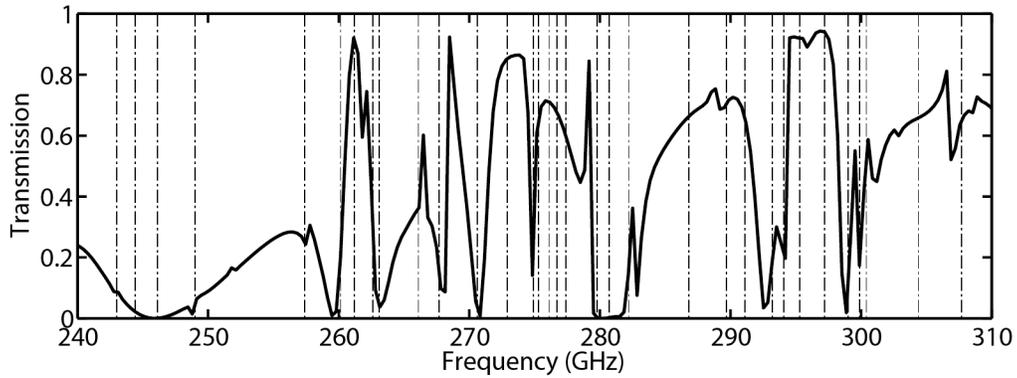

Fig. 7. Fiber transmission spectrum for fiber-metamaterial system with $N_l=10$ periods. Dotted vertical lines indicate spectral position of the Van Hove singularities as found from Fig. 2(b).

We now consider in more details formation of the resonant features in the fiber transmission spectrum when increasing the length of the fiber-metamaterial coupler. As an example, we consider changes in the line shape of a peak located at *~249 GHz*. In Fig. 8 we plot fiber transmission spectra for an increasing number of SSR periods. First, we note that the width of a resonance decreases rapidly when increasing the number of SRR periods from $N_l=1$ to $N_l=7$ (see Fig. 8(a)). Further increase in the number of SRR periods does not lead to a significant change in the peak width as seen in Fig. 8(b) where we present transmission spectra for $N_l=8-11$. This behavior is expected as the peak width corresponds to the coupling strength between the fundamental fiber mode (port mode) and a particular metamaterial mode. This strength is, in turn, proportional to the overlap integral between the fields of the two modes. When increasing the number of SRR periods, the modal fields of a finite-size fiber-metamaterial coupler converge to those of an infinitely periodic system. Therefore one expects that the value of coupling strength, and, hence, the transmission peak bandwidth should also converge to a certain finite value.

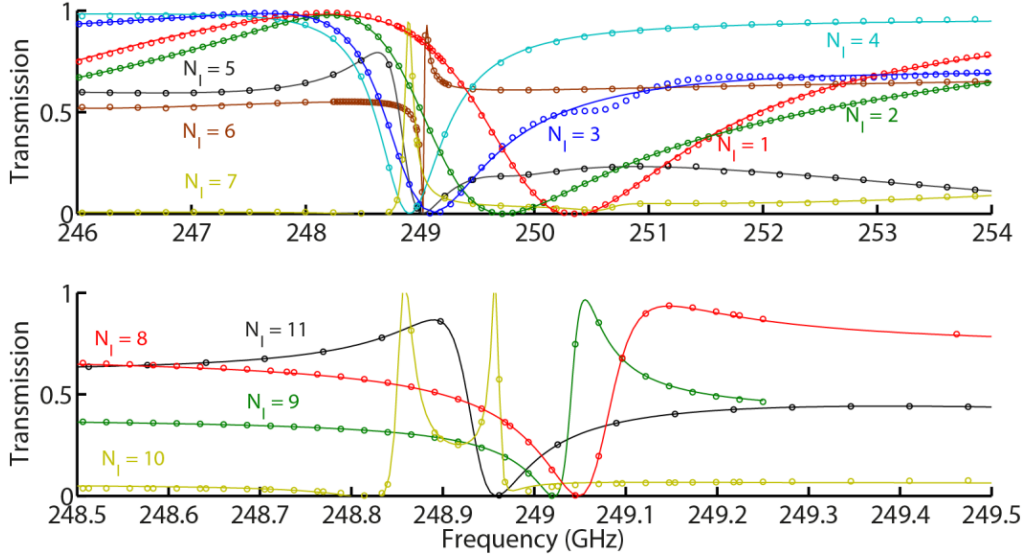

Fig. 8. Fiber transmission spectra for various numbers of SRR periods in the fiber-metamaterial coupler in the vicinity of a peak at ~249 GHz: (a) $N_l=1-7$. (b) $N_l=8-11$. Circles indicate numerical calculations, while solid lines are analytical fits using a single Fano line shape (13). (Exceptionally, data for $N_l=10$ is fitted using two Fano resonances).

Changes in the peak shapes can be studied quantitatively by fitting them with Fano line shapes and observing convergence of the Fano parameters when increasing the number of SRR periods. Particularly, using Fano line shapes similar to (11) we can extract the bandwidth and the asymmetry parameter for each of the peaks in Fig. 8. As the amplitudes of some transmission peaks in Fig. 8 are somewhat smaller than 1, we use the following generalized form of the Fano line shapes to fit the data:

$$|t|^2(f) = \frac{a}{1+q^2} \frac{(q\Gamma_\Phi + f - f_\Phi)^2}{\Gamma_\Phi^2 + (f - f_\Phi)^2} + c \ . \ (13)$$

where we restricted the fits to $\Gamma_\Phi > 0$, and $|q| < 1$. In Fig. 9 we show the results of the fitting and observe general convergence of the peak parameters when increasing the number of SRR periods. We note that one exception to our fitting procedure is the case of $N_l=10$. For this number of SRR periods, a second peak appears accidently in the vicinity of our main peak. In this case, the line shape (13) can no longer be used to perform the fit, therefore this point is omitted from Fig. 8. Unfortunately, calculations with the larger number of periods becomes impossible because of the virtual memory limitation (128G) of our machine.

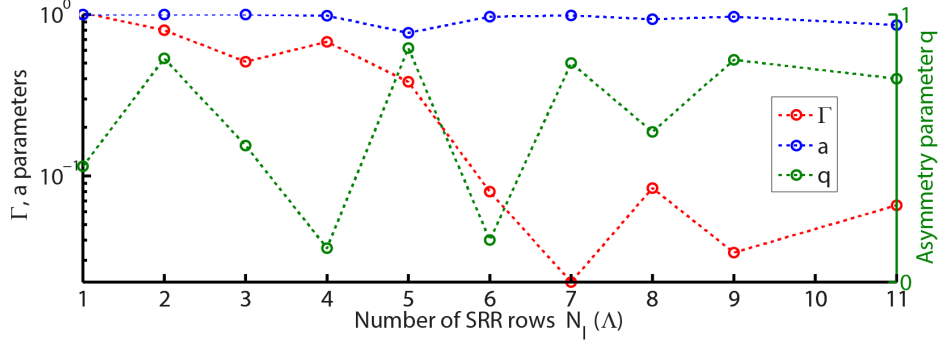

Fig. 9. Peak amplitude $a$ (blue), peak bandwidth $\Gamma_\Phi$ (red), and peak asymmetry $q$ (green) obtained from fitting the resonant line shape near 249 GHz (see Fig. 8) with Fano line shape.

## 8. Resonance engineering using band diagram calculations

It was shown in sections 5 and 6 that position and bandwidth of resonances in the fiber transmission spectrum can be tuned by varying separation between fiber and metamaterial. Frequency tuning using mechanical means, such as changing the fiber-metamaterial separation, is especially interesting for THz point-based devices due to ease of its practical implementation. As we have mentioned earlier, spectral position of the resonances in the fiber transmission spectrum can be predicted by identifying the frequencies of Van Hove singularities in the fiber-metamaterial band diagram. This method is considerably faster and less computationally intensive than a full transmission simulation. To ensure that Van Hove singularities found in the band diagram indeed manifest themselves in the fiber transmission spectrum, it is necessary to label all the optical states in the band diagram according to the relative amount of the electromagnetic fields in the fiber core. Then, a Van Hove singularity that is formed by avoiding crossing of modes with at least one of them having a significant presence in the fiber core will appear in the fiber transmission spectrum.

By varying the fiber-metamaterial separation and recomputing the band diagram for an infinite fiber-metamaterial system one can, therefore, tune position of the resonances in the fiber transmission spectrum, while avoiding a full transmission calculations.

## 9. Conclusion

In this work we studied the use of subwavelength terahertz fibers as light delivery and interrogation platform for probing metamaterials. Particularly, we studied transmission through a subwavelength fiber that is placed in parallel to a frequency selective surface. Thus defined coupler can operate both in the weak and strong coupling regimes depending on the field overlap between the fiber and metamaterial modes. This coupling is controlled conveniently by the fiber-metamaterial separation. Different polarization states can be realized in this arrangement with electric field either parallel or perpendicular to the metamaterial surface depending on polarization of the fiber mode.

Both the band diagram technique and the port-based scattering matrix technique were used to explain the nature of various resonances in the fiber transmission spectrum. We have concluded that spectral positions of most of the transmission peaks in the transmission simulation can be related to the positions of Van Hove singularities in the band diagram of a corresponding infinite periodic system. In the limit of weak coupling between the fiber and metamaterials modes, Van Hove singularities are found in the region of avoiding crossing between dispersion relation of a fiber mode and dispersion relation of a particular slab mode that was folded back into the first Brillouin zone due to its interaction with a periodic SRR array. Spectral shapes of most of the features in the fiber transmission spectrum were found to be of either Fano-type or EIT-type. Finally, we have demonstrated that center frequencies and bandwidths of these resonances and, as a consequence, spectral shapes of the corresponding transmission features can be efficiently tuned by varying the fiber-metamaterial separation.

We believe that this is the first time when interaction between THz subwavelength fiber and frequency selective surface was studied in details. Our main motivation behind this study is development of a convenient fiber-based platform for designing integrated THz devices based on fiber-metamaterial couplers for applications in sensing and signal processing.

## Acknowledgements

Funding for this project came from the NSERC strategic grant 430420 with the in-kind support from Honeywell Asca Inc., Vancouver Operations, and TeTechS Inc. Also, we would like to acknowledge Prof. M. Koch and his student N. Born for some fruitful discussions that we had with them at the beginning of the project in 2011. At that time Prof. M. Koch has contributed some experimental samples that we plan to use in our future experimental work.